\documentstyle[12pt]{article}
\input{epsf}
\newcommand{\be}{\begin{equation}}
\newcommand{\ee}{\end{equation}}

\newcommand{\R}{{\rm I \hspace{-0.52ex} R}}
\newcommand{\N}{{\rm I \hspace{-0.52ex} N}}
\title{Dirac oscillators and quasi-exactly solvable operators}
\author{{\large Y. Brihaye and A. Nininahazwe} \\
{\small Department of Mathematical Physics, University of Mons-Hainaut}\\
{\small Place du Parc, B-7000 Mons, Belgium }}
\date{\today}
\begin{document}
\begin{titlepage}
\maketitle
\begin{abstract}
The Dirac equation is considered in the background of
potentials of several types, namely
scalar and vector-potentials as well as
``Dirac-oscillator" potential or some of its generalisations.
We investigate the radial Dirac equation within a quite
general spherically symmetric form for these potentials
and we analyse some exactly and quasi exactly solvable
properties of the underlying matricial linear operators.
\end{abstract}
\end{titlepage}

\section{Introduction}

In quantum mechanics, the Schr\"odinger equations which can be
completely solved by algebraic methods are rather exceptional.
One can attenuate the condition of complete solvability
by asking that, at least, a few eigenvectors can be obtained
by an algebraic method.
The so called ``quasi exactly solvable"  (QES) equations
\cite{tur1,ush}  refer to a class of quantum hamiltonians
which possess precisely this property.
The corresponding operators can be set in correspondance
with finite dimensional representations of some Lie algebras and,
accordingly, a few of their eigenvectors can be computed by solving
an algebraic equation.

It turns out that QES Schr\"odinger equations often occur
as suitable extensions of exactly solvable ones.
The most famous example \cite{tur1} is the
one dimensional quantum harmonic oscillator.
When completed by a suitable quartic plus sextic potential
the equation of the harmonic
oscillator becomes QES.
In three dimensions, the prototype of exactly solvable
quantum hamiltonian corresponds to the charged particle in a central
Coulomb potential.
When the Coulomb potential is
supplemented by a confining potential of the form
$V=\omega^2r^2 + \lambda r$,
the equation becomes QES \cite{tur2};
the corresponding spectral
problem is, however, of type II \cite{tur1};
that is to say that the
energy eigenvalue of the initial Schr\"odinger problem is not
the spectral parameter of the algebraic equation.
A specific combination of the physical coupling
constants plays the role of the
spectral parameter of the QES equation; 
one energy level can be determined for the corresponding values
of the coupling constants.

The possibility of obtaining operators with interesting algebraic
properties can also be investigated for relativistic equations
like the Dirac equation which leads, in general, to a system of
coupled equations. Although the classification of QES equations
is well understood for scalar equations, this is not the case
for systems and a complete classification of matrix-valued 
QES  operators is still missing.
As a general rule the criteria to fullfill the QES property are
more severe for systems of equations and it turns out
interesting to consider ``type II" systems as well \cite{bk}.
The family of QES problems obtained in \cite{bk} was extended
in \cite{znojil} whom proposed a method to incorporate screened
Coulomb and scalar potentials.

More recently the Dirac-Pauli equation was reconsidered
\cite{roy} in the context of neutral particle interacting with
an electromagnetic field  and shown to
be strongly related to the case of a Dirac oscillator.
A similar problem was emphasized is 1+2 dimensions in
\cite{chiang0}. With suitable form of the radial potentials,
the corresponding Dirac equation can be quasi exactly solvable.

In this paper, we investigate the occurrence of explicit solutions
in the framework of the Dirac  equation coupled to
a class of external radial potentials, extending the cases of 
Dirac oscillator and the Dirac-Coulomb problem and generalizing
the choices of \cite{roy,chiang0}.
The general physical problem to deal with is presented in Sect.2.
Three cases for which the underlying operator is completely
solvable, namely the Dirac oscillator, the extended Dirac oscillator 
and the Dirac-Coulomb problem are presented in Sect.3; the emphasis 
is put on the construction
of equivalent operators preserving an infinite flag of vector spaces
of polynomials.
In both cases the quantization of the eigenenergy appears 
as a necessary condition for these invariant spaces to exist.
The general problem of constructing  QES operator out of the
generalized  Dirac oscillator  is adressed in Sect. 4

\section{Dirac equation and radial potentials}
We will study the radial equations
associated with the 3+1 dimensional Dirac equation
coupled to (radial) scalar and vector potentials. 
We start with the Dirac-Pauli equation~:
\be
\label{dirac}
 ( (i\partial_{\mu}-e A_{\mu}) \gamma^{\mu}
   - \frac{1}{2} \mu_n \sigma^{\mu \nu} F_{\mu \nu}
 - M - W(r) ) \psi(x_0,x_k) = 0
\ee
where $A_{\mu}$ is a vector potential, $W$ is a scalar potential
and  $F_{\mu \nu}$ is the electromagnetic field. The charge,
mass and anomalous magnetic moment of the spin-1/2 particle
are respectively denoted $e, M,\mu_n$.

\par
After standard manipulations, namely the separation
of the time variable and the separation of the angular
variables in a central potential, the radial equation
associated with (\ref{dirac}) takes a conventional form
of a 2*2 matrix equation \cite{roy}. Here we will assume the quite
general form
\be
\label{di1}
H    
\left(\begin{array}{c}
f\\
g
\end{array}\right)
\equiv
\left(\begin{array}{cc}
{d\over{dr}}-{\kappa\over r} + \mu_n (B-E) &M-\varepsilon-V+W\\
M+\varepsilon +V+W &{d\over{dr}}+{\kappa\over r} + \mu_n (B+E)
\end{array}\right)
\left(\begin{array}{c}
f\\
g
\end{array}\right)
= 0
\ee
where $\varepsilon$ is the energy parameter
while $\kappa$ is  the total angular momentum.
In this equation we have set four 
independent external radial potentials.
The parts $W(r)$ and $V(r)$ denote respectively
a  scalar potential and the time part of vector potential $A_{\mu}$.
The part $E(r)$ and $B(r)$ are related respectively
to a radial
electric and magnetic fields which couple 
through the anomalous
 momentum $\mu_n$ of the particle. In the following we will consider
 the mathematical problem of finding explicit solutions of the
 above equation  with the form of the generalized potentials  
 $E(r),B(r),V(r),W(r)$  fixed  {\it  a priori}  by hand.
 This general point of vue makes sense since e.g.
 the Dirac oscillator is recovered for
$\mu_n \vec E \rightarrow m \omega \vec r$.

In the following, we will restrict ourselves to radial
 potentials  of the forms
\be
\label{vw}
V(r)={\alpha\over r} + \sum^s_{i=1} \alpha_ir^i
\ \ \ , \ \ \ 
W(r)={\beta\over r} + \sum^s_{i=1} \beta_i r^i
\ \ \ , \ \ \
E(r) =  \sum^s_{i=0} \gamma_i r^i
\ee
along with \cite{roy} we assume $B(r)=0$.

\section{Exactly solvable cases}
In this section we construct explicitely the infinite flag
of invariant vector spaces for two particular cases for which the
operator above turns out to be exactly solvable. The technique
is such that the quantization of the energy comes out as a necessary
condition for the operator to preserve a vector space of the form
${\cal P}(n) \oplus {\cal P}(n-1)$, where ${\cal P}(k), k \in \N$ denotes
the polynomial of degree ot most $k$ in an appropriate  variable, say $x$.

Because  the vector space 
${\cal P}(n) \oplus {\cal P}(n-1)$
constitutes the basis of a particular representation of the 
graded-algebra osp(2,2) \cite{shif}, these results demonstrate
that the operator $H$ corresponding to these cases is indeed 
equivalent to an element of the envelopping algebra of osp(2,2).

\subsection{Dirac oscillator}
For this case one has   $V=0, W=0, E(r)= -r$.
The invariant spaces of the corresponding operator $H$ 
are revealed in terms  of spaces of polynomials
after we perform the following transformation
\be
   \tilde H = A^{-1} H A C
\ee
with the operator $A$ and constant matrix $C$ defined according to
\be
A=
e^{-\frac{\omega^2 r^2}{2}}\left(\begin{array}{cc}
r^{\theta} &0\\
0 &r^{\theta - 1}
\end{array}\right) \qquad , \qquad
C =
\left(\begin{array}{cc}
1 &\frac{2 \mu_n}{M + \varepsilon}\\
0 &1
\end{array}\right)
\ee
Choosing $\theta = \kappa$ and $\omega^2 = \mu_n$ 
and using  $x=r^2$ as a new variable, the Hamiltonian takes 
the form
\begin{equation}
\tilde H = x
\left(\begin{array}{cc}
0             &0\\
M+\varepsilon &0
\end{array}\right)
 + 2 x \frac{d}{dx}
\left(\begin{array}{cc}
1             &\frac{2 \mu_n}{M + \varepsilon}\\
0             &1
\end{array}\right)
+
\left(\begin{array}{cc}
0             &{M - \varepsilon}\\
0             &2 \kappa-1
\end{array}\right)
\end{equation}
which turns out to be a linear combination of the generators
of the super algebra osp(2,2) 
(in the suitable representation first pointed out in \cite{shif}),
provided  the energy parameter
$\varepsilon$ is of the form
\be
   \varepsilon^2 = M^2 + 4 n \mu_n \ \ \ , \ \ \ n \in \N
\ee
reproducing the spectrum of the Dirac oscillator equation 
(see e.g. \cite{lin}).
Indeed, in this case, the matrix element $\tilde H_{12}$
becomes proportional to the operator $x \frac{d}{dx}-n$ and
correspondingly, the operator preserves the vector space
of polynomials of the
form $(p_{n-1}(x), p_{n}(x))^t$.

This way of obtaining the spectrum of the Dirac operator
by enforcing it to preserve an infinite family of (finite dimensional)
vector spaces
 reveals the hidden algebraic structure of the Dirac
oscillator equation by its relation with osp(2,2).

\subsection{Extended Dirac oscillator}
 The result of the standard Dirac oscillator discussed
 above can be extended to a case of type II exactly solvable
 system if the various potentials are suitably chosen.
 In the purpose to exhibit a such possible extension, let us
 consider the potentials with the following form~:
 \be
 \label{potesii}
V(r) = {\alpha \over r} \qquad ,
\qquad W(r) = \beta_1 r  + {\beta \over r} \qquad ,   
\qquad E(r) = \gamma_0 + \gamma_1 r
 \ \ .
 \ee
In order to reveal the hidden algebra structure of the corresponding
$H$ several changes of variables and/or functions  are necessary.
First, it is convenient to write $\gamma_1 = R \cos(2 \omega)$,
$\beta_1 = R \sin(2 \omega)$.Secondly, we "gauge rotate" the operator
by means of
\be
          \  U  \ S \exp(-R r^2/2 - T r)
\ \ , \ \
S = {\rm diag}( r^{\theta}, r^{\tilde \theta})   \ \ , \ \
  U=
\left(\begin{array}{cc}
\cos \omega & -\sin \omega \\
\sin \omega & \cos \omega
\end{array}\right)
\ee
Then, the following values have to be imposed
(with  $c \equiv \cos(2\omega)$ , $s \equiv \sin(2 \omega)$)
\be
   \alpha = 0 \  , \ \beta = - \kappa \tan{2\omega} \ \ , \
   \gamma_0 = - M \tan{2 \omega} \ \ , T = 0   \ , \ \theta = \kappa c \ \ , \  \tilde \theta-\theta = \Delta = -1
\ee
and if the so obtained operator, say $\tilde H$, is further
transformed by means of
 \be
      \hat H = \tilde H   
      \left(\begin{array}{cc}
1 & y    \\
0 & 1
\end{array}\right)   \ \ , \ \
 y \equiv \frac{2 R }{\varepsilon + \frac{M}{c} }
 \ee
and the variable $x = r^2$ is used,
we finally obtain an operator which preserves 
$(p_{n-1}(x), p_{n}(x))^t$.
The energy quantization formula reads in this case
\be
   E^2 = \frac {M^2}{c^2} + 2 n R \ \ , \ \ n \in \R
\ee
Finally, the quantum Hamiltonian obtained in this way
generalises the conventional
Dirac oscillator by means of the addition of the angle $\omega$
appearing through the coupling constants $\beta_1,\gamma_1$.
All the other coupling constants (namely $\gamma_0, \beta, \alpha$) 
are fixed in terms of $M, \kappa$ and $\omega$.
Since $\alpha = 0$ turns out to be
a necessary condition, the  attempt above further
reveals the difficulty
(if not the impossibility...) to obtain a hidden algebraic
form of the combined relativistic problem
Dirac Oscillator + Coulomb potential.
In the following section, we will succeed in producing
partial algebraic solutions to this problem.

\subsection{Dirac-Coulomb problem}
This case is well known and was presented in \cite{bk}
adopting the point of view of exactly solvable operators.
However
we present briefly the construction here for the purpose 
of completeness and for the sake of comparaison
with the case studied in the previous section.
The conditions on the potentials are
 $\alpha_i = \beta_i = 0$, $E=B=0$. 
The  relevant transformation of the Hamiltonian reads
 \be
   \tilde H = A U^{-1} H U C
\ee
with
\be
U =
r^{\theta} {\rm exp}(-\lambda r) \left(\begin{array}{cc}
u_- &u_-\\
u_+ &-u_+
\end{array}\right)
\qquad , \qquad u_{\pm} \equiv \sqrt{M \pm \varepsilon}
\ee
\be
A=
\left(\begin{array}{cc}
1 &0\\
-1 &1
\end{array}\right) \qquad , \qquad C =
\left(\begin{array}{cc}
1 &-1\\
1 &0
\end{array}\right)
\ee

Choosing the arbitrary parameters appearing in 
the operator $U$ according to
\be
\theta^2=\kappa^2+\beta^2-\alpha^2 \ \ \ , \ \ \
\lambda^2=M^2-{\varepsilon}^2
\ee
and multiplying the equations by $r$, 
leads to the new operator $\tilde H$ which reads
(up to diagonal elements which depend
on  $r{d\over dr}$)
\be
\label{opd}
\tilde H = -2\sqrt{u_+u_-}
\left(\begin{array}{cc}
0 &0\\
r &0
\end{array}\right)
- 2\alpha \sqrt{u_-\over{u_+}}
\left(\begin{array} {cc}
0 &0\\
1&0
\end{array}\right)
-
\left(\begin{array}{cc}
0 & r{d\over {dr}} - \eta\\
0 &0
\end{array}\right)
\ee
with
\be
\eta \equiv {\alpha\varepsilon-M\beta\over{\sqrt{M^2-{\varepsilon}^2}}}
- \sqrt{\kappa^2+\beta^2
-\alpha^2}
\ee

If the quantity $\eta$ is imposed to be an integer,
$n$, the operator $\tilde H$
manifestly preserves the vector space $P(n-1) \oplus P(n)$ 
for $n=1,2,3,\dots$
(as before $P(n)$ denotes the space of polynomials of degree
less or equal to $n$ in $r$).

The condition $\eta=n$ leads to
the quantization of the energy of the system
\cite{villalba,torres}.
The celebrated spectrum of
the relativistic hydrogen atom
(in the case $\beta=0$). As for the case of the Dirac oscillator
it is obtained by requiring that the reduced spherically
symmetric Hamiltonian can be written in terms of the
generators \cite{shif} of the super algebra osp(2,2).
The infinite series of
finite dimensional vector spaces preserved by this
realization corresponds to the eigenvector of the problem.

\section{Quasi exactly solvable cases}
\subsection{Planar case}
Here we consider the case of a planar Dirac electron 
in Coulomb plus magnetic field.
This case was studied in   \cite{chiang0,chiang}. In the general
framework of Eq.(\ref{di1}) it corresponds to $B(r)=0$, $W(r)=0$
and $E(r) \rightarrow e \tilde B r/2$. We also note two main
differences that $r$ should now be interpreted as
a two-dimensional radial variable and that $\kappa$ now
takes half-integer values. Also, in the following discussion
we will rescale the radial variable according to $x = r\sqrt{e\tilde B}$.
Posing, along with \cite{chiang0}, $f= x^{\gamma} exp{(-x^2/4)}  Q(x)$,
$g= x^{\gamma} exp{(-x^2/4)}  P(x)$
where $P,Q$ are polynomials leading to
 the following equations for $P,Q$~:
\begin{eqnarray}
&P' + \frac{\kappa + \gamma}{x} P - (\varepsilon-M+ \frac{\alpha}{x})Q &= 0 \ \ , \ \
\\
&Q' + \frac{\gamma-\kappa}{x} Q - x Q + (\varepsilon +M+ \frac{\alpha}{x})P &= 0
\end{eqnarray}
Where $\varepsilon$ and $M$ have been rescaled appropriately.
The polynomials $Q,P$ should have degrees $n, n+1$ respectively,
nevertheless the number of equations $2n+5$ exceeds by one unit
the number of free parameters $2n+4$ and therefore a algebraic
solution only exist for specific values of $\alpha$. Two of the
equations allow to
fix immediately the values of $\gamma$ and $\varepsilon$ according to
\begin{equation}
\gamma = \sqrt{\kappa^2 - \alpha^2} \ \ , \ \
{\varepsilon}^2 = M^2 + \gamma + \kappa + n + 1
\end{equation}
Unfortunately, we did not manage
to find the extra condition for $\alpha$ in a simple way for generic
values of $\kappa$, $M$ and $n$.

In the simplest case, $n=0$, we find
\begin{equation}
\varepsilon = -\frac{1}{2} (M \pm \sqrt{M^2 + 2}) \ \ , \ \
\alpha^2 = - \frac{1+8\kappa {\varepsilon}^2}{16 {\varepsilon}^4}
\end{equation}
showing that two branches of values of $\alpha$ lead
to polynomial solutions.  However these eigenvectors are
available only for $\kappa < 0$.
In particular, this indicates that
the  cylindrical symmetric ground state
is not one of the algebraic solutions.

Already for $n=1$ the final equation relating $\alpha, M , \kappa$
is very involved.
The allowed values of $\alpha$ with $M,\kappa$ fixed
can be determined numerically. 
For the case $\kappa = 1/2$, there are four possible values of 
$\alpha$ and two corresponding values of the energy $\varepsilon$.
They are presented on Fig.1 for $M \in [0,1]$.   For $\kappa = 3/2$
the picture is qualitatively similar to Fig.1.

\subsection{Extended Dirac oscillators}
The case treated in the previous section is physically 
important, but we have seen that no algebraic solution
can be constructed for generic values of the Coulomb
and oscillator coupling constants, respectively $\alpha, \tilde B$.
We would like to construct a model where algebraic solutions
exist for generic values of the Coulomb and oscillator constants.
In this purpose,
we consider Eq.(\ref{di1}) with an extended choice for
oscillator's parameters. Namely, we set
\be
\label{potcc}
V(r) = {\alpha \over r} \qquad ,
\qquad W(r) = \beta_1 r    \qquad E(r) = \gamma_0 + \gamma_1 r
 \ \ .
\ee
This choice constitutes one possible relativistic generalisation of the
non-relativistic problem considered in \cite{tur2}
Here the Coulomb interaction is supplemented by two types of
confining (i.e. linear in $r$) interactions.
Note that, adding a constant to the potential $W(r)$
is equivalent to redefine the mass $M$. To the contrary,
the constant $\gamma_0$ cannot be eliminated by a redefinition
of the physical quantities.

We look for solutions of the form
\be
\label{wavfun}
\left(\begin{array}{c}
f\\
g
\end{array}\right)
= r^{\theta} \exp (-{\lambda_2\over 2} r^2 - \lambda_1 r)
\left(\begin{array}{c}
\tilde p\\
\tilde q
\end{array}\right)
\ee
where $\theta, \lambda_1, \lambda_2$ are constants and $\tilde p(r),\tilde q(r)$
are chosen as polynomials in $r$.
Inserting (\ref{wavfun}) into (\ref{di1}), we obtain the counterpart of
(\ref{opd}) with
\be
\tilde H =
\left(\begin{array}{cc}
\label{opedi}
  D -\lambda_2 r^2 - \lambda_1 r + \theta -\kappa + \gamma_0 r + \gamma_1 r^2
& \beta_1 r^2 - \alpha + r(M-\varepsilon)\\
  \beta_1 r^2 + \alpha + r(M+\varepsilon)
& D -\lambda_2 r^2 - \lambda_1 r + \theta + \kappa - \gamma_0 r - \gamma_1 r^2
\end{array}\right)
\ee
where $D$ is the dilatation operator $D = r \frac{d}{dr}$.
Transforming the system according to
\be
U^{-1} \tilde H U
\left(\begin{array}{c}
\tilde p\\
\tilde q
\end{array}\right)
= 0
 \ \ ,  \ \
 U=
\left(\begin{array}{cc}
\cos \omega & -\sin \omega \\
\sin \omega & \cos \omega
\end{array}\right)
\ee
with $R =\sqrt{\gamma_1^2 + \beta_1^2}$,
$\omega = (1/2)\arctan(\beta_1/\gamma_1)$.
we obtain  the following conditions for polynomial solutions to exist:
\be
\lambda_2 = R \quad , \quad
\lambda_1 = \frac{\beta_1 M + \gamma_0 \gamma_1}{R}
\ee
with
\be
\label{deg}
{\rm deg}\ \tilde p = n \quad, \quad {\rm deg}\ \tilde q = n-1 \ .
\ee
The final equations for $\tilde p, \tilde q$ then read
\begin{eqnarray}
\label{equ1}
&(D+A_2)\tilde p + (A_1r+A_0)\tilde q &= 0\nonumber \\
&(D+C_2r^2+C_1r+C_0) \tilde q +(D_1r+D_0)\tilde p &= 0
\end{eqnarray}
where
\begin{eqnarray}
&A_2 &= \theta - \frac{\gamma_1 \kappa}{R}\nonumber \\
&A_1 &= \gamma_1^2 \frac{(\varepsilon - M)}{\beta_1 R}
+ \gamma_1 \frac{(\beta_1 \gamma_0 + \varepsilon R - M R)}{\beta_1 R}
+ \frac{\beta_1 \varepsilon + \gamma_0 R}{R}\nonumber \\
& A_0 &= \gamma_1^2 \frac{\alpha}{\beta_1 R}
+ \gamma_1 \frac{(\alpha R - \beta_1 \kappa)}{\beta_1 R}
+ \frac{\alpha \beta_1 - \kappa R}{R}\nonumber \\
& C_2 &= -2R \nonumber \\
& C_1 &= -2\frac{\beta_1 M + \gamma_0 \gamma_1}{R}\nonumber \\
& C_0 &= \frac{\gamma_1 \kappa + \theta R}{R}\nonumber \\
& D_1 &=-(\gamma_1^2 \frac{M - \varepsilon}{\beta_1 R}
   + \gamma_1 \frac{\varepsilon R - M R - \beta_1 \gamma_0}{\beta_1 R}
   + \frac{\gamma_0 R - \beta_1 \varepsilon}{R}
   )\nonumber \\
& D_0 &= -\gamma_1^2 \frac{\alpha}{\beta_1 R}
+ \gamma_1 \frac{\alpha R
+ \beta_1 \kappa}{\beta_1 R}
+  \frac{\kappa R - \alpha \beta_1}{R}
\end{eqnarray}

Eq. (\ref{equ1}) with (\ref{deg}) leads to a set of $2n+3$ algebraic
equations in $2n+2$ parameters
($\theta, E$ and $2n$ coefficients for $\tilde p$ and $\tilde q$); so
it cannot be solved  for generic
values of the physical coupling constants
$\alpha, \beta_1, \gamma_0, \gamma_1, M$. However,
considering one of these physical parameters  as free
(e.g. the parameter $\gamma_0$)
we are lead to consistent equations which, in principle, possess solutions
and fix one energy level $\varepsilon$ and the constant
$\gamma_0$ as functions of
the parameters $\alpha, \beta_1, \gamma_1, M$ and of $\kappa$.

The analysis of terms of highest and lowest degrees
in   $r$  in these linear equations lead respectively
to the following condition for the eigenvectors
\be
\label{solu}
        \tilde p = x^n + o(x^{n-1}) \ \ \ , \ \ \
        \tilde q = - \frac{n+A_2}{A_1} x^{n-1} + o(x^{n-2}).
\ee
\be
\label{cond}
       \varepsilon^2 = 2(R(n+\theta)-\gamma_1 \kappa)
       + \frac{M^2 \gamma_1^2 + \beta_1^2 \gamma_0^2
       - 2 M \beta_1 \gamma_0 \gamma_1}{R^2} \ \ \ , \ \ \
       \theta = \sqrt{\kappa^2 - \alpha^2} \quad , \quad
       \kappa = \pm 1 , \pm 2 , \dots
\ee
The first of these equations can be used to determine $\varepsilon$
while the second allows one to determine the parameter $\theta$.

We have analysed the equations numerically for the case
$n=1, \kappa = 1$ corresponding to the ground state of the
equation. Fixing symbolically $\gamma_1 =3$,$\beta_1=4$ and $M=1$
we found that two possible values occur for the parameter
$\gamma_0$. These values are real for $\alpha \in [0.097, 0.92]$
and become complex outside that interval. The two branches of
values are presented on Fig. 2.

The results obtained above suggest that the reduced  spherically
symmetric Hamiltonian is equivalent to an operator preserving the
vector space $P(n-1) \oplus P(n)$. However, in spite of our efforts,
we could not find a change of basis making this job and we believe 
that it is not. Nevertheless, if we combine the two equations in order to
obtain decoupled second order equations for the two components
of the spinor, we got linear operators of the form \cite{chiang}
\begin{equation}
T \equiv (x^2 + x_0 x) \frac{d^2}{dx^2}
+ (-x^2(x+x_0) + 2\beta(x+x_0)) \frac{d}{dx}
+ \varepsilon x(x+x_0) + (b-c)x + b x_0
\end{equation}
where
\begin{equation}
\tilde \varepsilon = 
\varepsilon^2 - M^2 - \kappa - \gamma -1 \ \ , \ \
b = 2 \varepsilon \alpha + (\kappa - \gamma)\frac{\varepsilon+M}{\alpha} \ \ , \ \
c = \frac{\alpha}{\varepsilon+M} + (\kappa- \gamma) 
\frac{\varepsilon+M}{\alpha}
\end{equation}
A necessary condition for the equation $T Q_n(x) = 0$ to possess
polynomial solution of degree $n$ in $x$ requires 
$\tilde \varepsilon = n$,
quantizing the possible values of the energy. Once this is fixed,
the operator can be set in the form
\begin{equation}
T = -x^2(x \frac{d}{dx} - n) -x_0 x (x \frac{d}{dx}-n) + (b-c)x + S_{QES}
=  x T_{QES} + S_{QES}
\ \
\end{equation}
where $T_{QES}, S_{QES}$ can be expressed in terms of
the three basic generators $J^+_n = x(x \frac{d}{dx}-n)$, 
$J^0_n = (x \frac{d}{dx}- \frac{n}{2})$ , $J^-_n = \frac{d}{dx}$.
This provides an alternative demonstation that $T$ can be
expressed as an element of the envelopping algebra of
osp(2,2), as pointed out recently in \cite{chiang0}.
Once set in this form, the equation $T Q_n(x) = 0$ leads
to a system of $n+2$ linear equations in $n$ parameters
($n$ parameters in $Q_n$ , remember that the parameter
$\gamma$ is fixed by $\gamma = \sqrt{\kappa^2 - \alpha^2}$).

So naively, one would expect the equations to fix two
conditions among the coupling constants of the model,
contrasting with the counting of equations and parameters
done with the systems of first order equations.

However, due to the fact that the second order equation
is obtained from the first order ones, it turns out that
the two extra conditions are indeed consistent with each other
and there is effectively {\bf one} extra relation.
To our knowledge, this property is not at all apparent by
just looking at the second order equation. 
This construction can nevertheless be used to extend the class
of one dimensional quasi exactly solvable operators away
from the class of operators which are directly expressible
in the envelopping algebra of sl(2,R).


\section{Outlook}
\par We have shown that algebraic solutions of the
Dirac + Coulomb + confining potential equations
(with both Dirac oscillator and normal harmonic potentials)
can admit some explicit bound states.
A generalisation of the Dirac oscillator has been obtained
which can be completely solved algebraically
(i.e. it is exactly solvable) and posseses a hidden algebra
related to osp(2,2).

The mixed case, with both oscillators and Coulomb-terms,
leads generally to an overdetermined systems of equations but,
allowing one of the coupling constants to be ``free", leads to
a sufficient number
parameters and the system of equations can be solved consistently
by means of algebraic techniques. 
We believe that our results could be generalized
by including screened potentials and using the ideas of \cite{znojil}.

\begin{figure}
\epsfysize=22cm
\epsffile{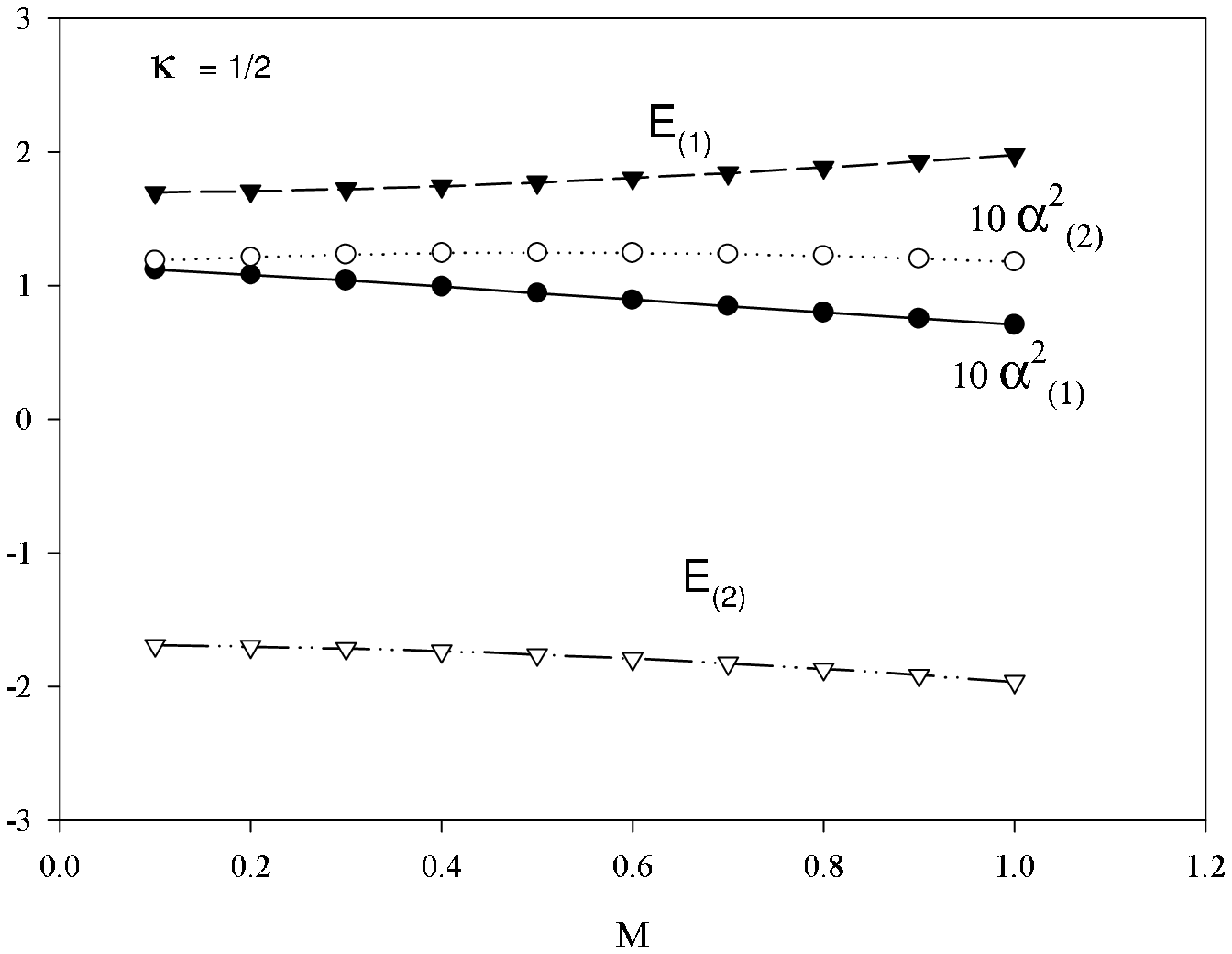}
\vskip -3cm
\caption{\label{Fig.1} The values of $\alpha$ and of the
energy are plotted as functions of M in the case $\kappa = 1/2$.
}
\end{figure}
\begin{figure}
\epsfysize=22cm
\epsffile{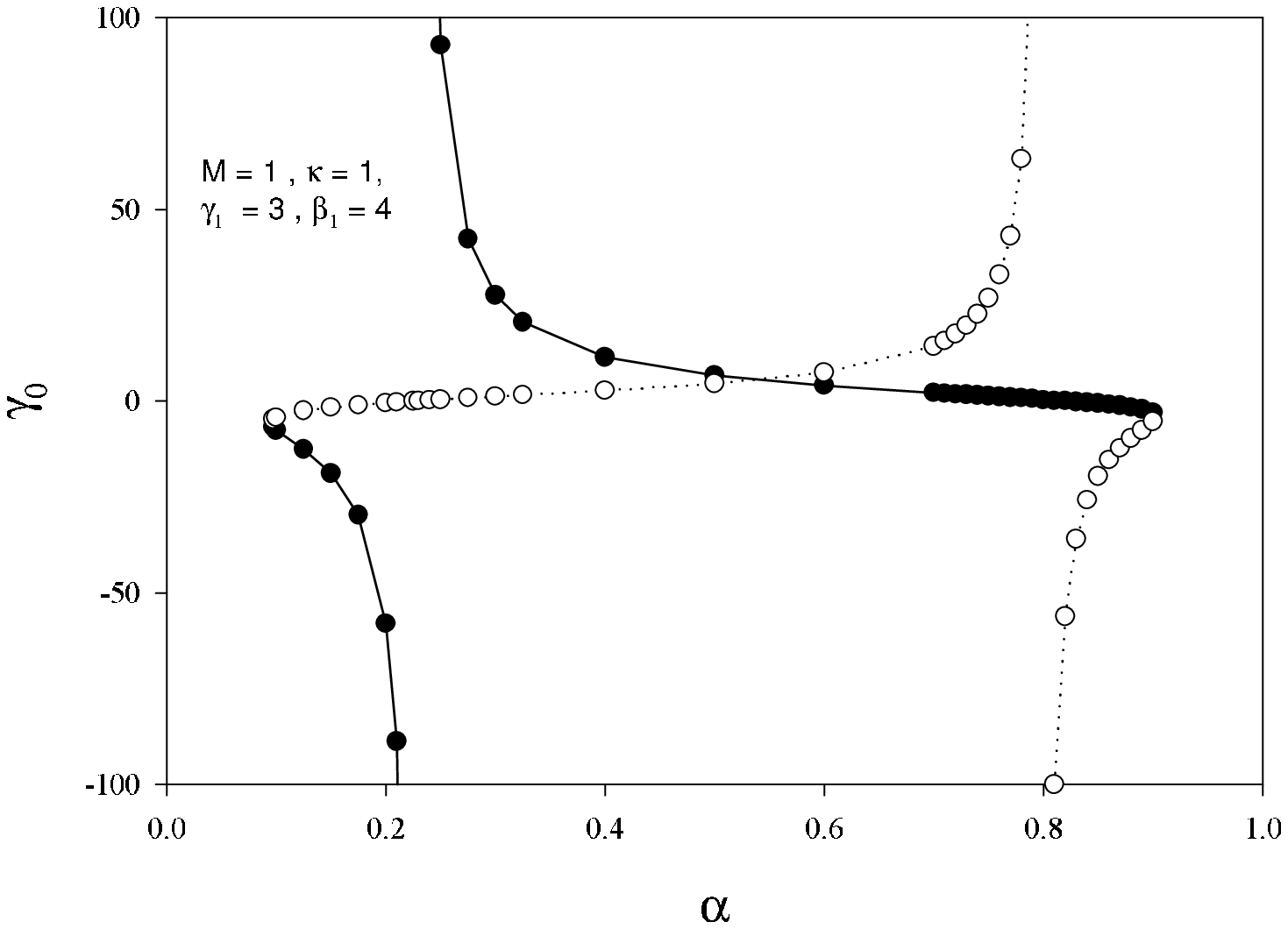}
\vskip -3cm
\caption{\label{Fig.2} The two possibles values of the parameter
$\gamma_0$ are plotted as functions of the Coulomb coupling constant
$\alpha$ for the "ground state`` solution corresponding to
mixed oscillators.
}
\end{figure}


\newpage
\begin{thebibliography}{6}
\bibitem{tur1} A.V. Turbiner, Comm. Math. Phys.{\bf{118}}, 467 (1988).
\bibitem{ush} A.G.Ushveridze, "{\it Quasi exact solvability in 
quantum mechanics}", Institute of Physics Publishing, Bristol
and Philadelphia (1993).
\bibitem{tur2} A.V. Turbiner, Phys. Rev. {\bf{A50}}, 5335 (1994).
\bibitem{bk} Y. Brihaye,  P. Kosinski,  Mod. Phys. Lett. {\bf A 14}, 
2579 (1999).
\bibitem{znojil}M.Znojil, Mod. Phys. Lett. {\bf A 14}, 863 (1999).
\bibitem{roy} C.-L. Ho and P. Roy, Annals of Physics {\bf 312},
161 (2004)
\bibitem{chiang0} C.-M. Chiang and C.-L Ho,
Quasi-Exact Solvability of Planar Dirac Electron in Coulomb
and Magnetic Fields, quant-ph/0501035.
\bibitem{shif} M. V. Shifman and A. V. Turbiner, Commun. Math. Phys.
{\bf 126} (1989) 347.
\bibitem{villalba} V. Villalba, J. Math. Phys. {\bf{36}}, 3332 (1995).
\bibitem{torres} G. Torres del Castillo, L. Cortes-Curantli, J. Math.
Phys. {\bf 38} (1997) 2996.
\bibitem{chiang} C.-M. Chiang and C.-L Ho, J. Math. Phys. {\bf 43}
(2002) 43.
\bibitem{lin} Q.-G. Lin, J. Phys. {\bf G 25}, 1795 (1999).
\end{thebibliography}
\end{document}